# Optimization of scandium oxide growth by high pressure sputtering on silicon

P.C. Feijoo, M.A. Pampillón, E. San Andrés, M.L. Lucía

Dpto. Física Aplicada III (Electricidad y Electrónica), Universidad Complutense de Madrid, 28040, Madrid, Spain

**Abstract**

This work demonstrates the viability of scandium oxide deposition on silicon by means of high pressure sputtering. Deposition pressure and radio frequency power are varied for optimization of the properties of the thin films and the $ScO_x$/Si interface. The physical characterization was performed by ellipsometry, Fourier transform infrared spectroscopy, x-ray diffraction and transmission electron microscopy. Aluminum gate electrodes were evaporated for metal–insulator–semiconductor (MIS) fabrication. From the electrical characterization of the MIS devices, the density of interfacial defects is found to decrease with deposition pressure, showing a reduced plasma damage of the substrate surface for higher pressures. This is also supported by lower flatband voltage shifts in the capacitance versus voltage hysteresis curves. Sputtering at high pressures (above 100 Pa) reduces the interfacial $SiO_x$ formation, according to the infrared spectra. The growth rates decrease with deposition pressure, so a very accurate control of the layer thicknesses could be provided.

1. **Introduction**

Scandium oxide can become an important material in the next generation of silicon transistors and memory devices. This material has a band gap of 6 eV [1] and a relative permittivity of 13 [2] which makes it suitable for flash memory applications [3]. It can be used for electrical isolation between the control gate and the charge storage layer for both floating gate memories (flotox) and charge trap memories. In flotox memories, the insulator is sandwiched between the poly-Si gates [4] and is called inter-poly-Si dielectric. In charge trap memories, the insulator or blocking oxide lies in contact with a silicon nitride trapping layer and a p-type metal [5]. Also, ternary based compounds such as $GdScO_3$, $DyScO_3$ and $LaScO_3$ have attracted interest in CMOS circuits because of their high permittivity and good stability in contact with Si [6–8].

In addition to this, $Sc_2O_3$ thin films find applications in several fields like e-beam lithography resist [9], damage resistant layer [10], antireflection coatings in light emitting diodes and high power ultraviolet lasers [11, 12], constituent of ferroelectric thin films [13], etc.

Currently, $Sc_2O_3$ thin films with good properties are deposited by atomic layer deposition (ALD) [14, 15]. The present work studies the physical and interfacial properties of scandium oxide films deposited by high pressure sputtering (HPS). This system was demonstrated to be adequate for other high κ dielectric growth [16–18]. During the deposition process, the pressure is maintained in the 20–200 Pa range, about three orders of magnitude higher than in a conventional sputtering system. Thus, the mean free path of the sputtered species in the plasma was very short. This means that the atoms reach the substrate with very low energy (~$k_BT$, $k_B$ being the Boltzmann constant and T the temperature) by a purely diffusion process. The low energy of the arriving species prevents the damage of the substrate and the growing film itself.

In this article, deposition of $ScO_x$ films by HPS on silicon is characterized for different growing conditions, with the aim of optimizing the properties of the films and the $ScO_x$/Si interface. The morphology, composition and crystal structure of the films are investigated by means of transmission electron microscopy (TEM), glancing incidence x-ray diffraction (GIXRD) and Fourier transform infrared spectroscopy (FTIR). The thickness is determined by ellipsometry. Metal– insulator– semiconductor (MIS) devices were fabricated to analyze the electrical properties of $ScO_x$.

## 2. Experimental details

Scandium oxide films were grown on two different kinds of substrates. For structural characterization, double side polished 2 inch Si wafers with a resistivity of 200–1000 Ω cm were used. For electrical characterization, metal/ScO$_x$ stacks were deposited on single side polished 2 inch Si wafers with a resistivity of 1.5–5.0 Ω cm. In both cases the wafers were n-type doped with phosphorus (n-type Si) and the top surface corresponded to (1 0 0) plane. The HPS system used in this work guarantees uniformity of the grown films in an area of 1×1 cm$^2$, although it could be easily modified in order to provide uniform layers over larger areas.

ScO$_x$ layers were deposited by HPS in a pure Ar atmosphere, using several conditions of pressure and radio frequency (rf) power. Prior to deposition, the base pressure was around 1–2×10$^{-4}$ Pa while the pressures of the processes ranged from 25 to 130 Pa. A commercial 4.5 cm diameter 99.99% pure Sc$_2$O$_3$ target was used (impurities consist of 0.003% of Ca and lower traces of Cd, Cr and Pd). The 13.54 MHz rf power was changed from 30 to 50 W. Depositions were performed for 30 min maintaining a constant substrate temperature of 200 °C. In order to get a thicker layer for structural characterization, one sample was fabricated at 50 Pa and 40 W for 2 h.

MIS devices of areas ranging from 50×50 to 630×630 μm$^2$ were fabricated on the low resistivity substrates. The gate electrode square openings were defined with a lithography process (negative photoresist n-LOF 2070). Afterwards, 130 nm thick Al metal electrodes were deposited by e-beam evaporation, followed by the lift-off of the photoresist. A 150 nm thick Al layer was evaporated on the backside to obtain the substrate ohmic contact. After fabrication, the samples were annealed in forming gas (FGA) for 20 min at 450 °C for the passivation of interfacial defects.

The emission lines of the HPS glow discharge were registered by a Jobin Yvon H-25 monochromator attached to a photon counting system, measuring the plasma emission at wavelengths between 280 and 520 nm, with a resolution of 0.1 nm. In order to analyze the bonding structure of the films and the interfaces, FTIR spectra of the samples were measured in the 4000–400 cm$^{-1}$ region using a Nicolet Magna-IR 750 series II spectrometer working in transmission mode at normal incidence. These spectra were corrected by subtracting the spectrum of a bare Si substrate of the same lot to remove the substrate absorbance. To avoid the native oxide presence on the reference spectrum, the Si substrate was immersed in HF 1:50 for 30 s immediately before the FTIR measurement. Thicknesses of the films were extracted from ellipsometry, with a Nanofilm EP$^3$ ellipsometer. GIXRD was measured by an XPERT MRD of Panalytical at $\omega=0.5°$ for crystalline structure identification. The wavelength of the x-ray beam was 1.541 Å, corresponding to the Cu K$\alpha$ line.

The samples were also characterized by means of cross sectional TEM using a JEOL-JEM-2000FX at 200 keV. A cross sectional TEM sample was prepared from the thicker ScO$_x$ layer (deposited for 2 h): two slices of the sample were cut and pasted so that the ScO$_x$ films faced each other; the stack was polished mechanically until a thickness of 100 μm; a dimple-grinder dug a sphere-shaped dip; and finally, an Ar ion beam milled the sample until a hole was opened. For the MIS structures, TEM samples were prepared by the H-bar focused ion beam technique [19]: the sample was thinned to 70 μm and an Ar ion beam cut two deep trenches to leave a lamella. In both kinds of TEM sample preparation, the more harmful step is the latter. However the damage is minimal since both processes are optimized for Si based samples.

The capacitance and conductance of the MIS capacitors were measured as a function of the gate bias voltage by an Agilent 4294A impedance analyzer. Finally, hysteresis curves were taken and related to the densities of interfacial defects D$_{it}$, estimated by the conductance method [20].

3. **Results and discussion**

*3.1. Structural characterization*

In the first place, the structural characterization of the thick $ScO_x$ film (deposited over 2 h) will be discussed. In Fig. 1, a cross sectional TEM image of this sample is shown. It can be noted that these conditions produce an ~34 nm thick $ScO_x$ layer. Electron diffraction fringes indicate a polycrystalline character of the film. However, nanocrystals are not so clearly observed in the region close to the $ScO_x$/Si interface, suggesting an amorphous initial growth stage. On the other hand, as is usual with other high κ dielectrics deposited on Si, such as $HfO_2$ or $TiO_2$, a $SiO_x$-like interfacial layer of 1.7±0.3 nm appears between the high κ and the Si.

The GIXRD pattern of the sample is represented in Fig. 2. The most important peaks of the $Sc_2O_3$ cubic bixbyite phase can be clearly identified. This is the only known polymorph phase of the stoichiometric scandium oxide [21]. No further peaks are found, an indication that this is the only crystalline phase. The agreement of the obtained pattern (peak position and relative intensities) with the typical diffraction pattern of the cubic bixbyite $Sc_2O_3$ powder suggests that no preferential direction in the polycrystalline growth is observed. This is also supported by the TEM image in Fig. 1, but it contrasts with previous $HfO_2$ results fabricated with the same HPS system [22], which showed a columnar growth.

The FTIR spectrum of the sample between 1200 $cm^{-1}$ and 400 $cm^{-1}$ is depicted in Fig. 3(a). The higher wavenumbers do not show any relevant peak, and they have been omitted for the sake of clarity. In the figure, a peak at 667 $cm^{-1}$ is found, which is related to residual $CO_2$ present in the camera, and two more at about 730 and 610 $cm^{-1}$, due to the absorption of a Si phonon [23]. These two peaks can be attributed to a small wafer thickness difference between the sample and reference substrates. Thus, these peaks will not be further discussed. A broad band can be observed at around 1070 $cm^{-1}$ that corresponds with the Si–O stretching vibration mode [24]. On the other hand, the increase in the absorbance in the region around 600–400 $cm^{-1}$ can be related to the Sc\O bond, as literature shows [13, 25] and as can be concluded from comparison with the measured spectrum of $Sc_2O_3$ powder, also shown in Fig. 3(b). From these results, it is confirmed that a polycrystalline $ScO_x$ layer is deposited on Si, and an interfacial layer of $SiO_x$ grows from the reaction with the substrate.

Now, the dependence of the properties of the films with the deposition conditions will be explored. The thickness measured from ellipsometry of the $ScO_x$ layers as a function of the rf power is depicted in Fig. 4. As it could be expected, higher powers mean thicker layers due to the increase in the energy of the particles in the plasma which collide with the target. Then, the amount of sputtered atoms is increased and so is the growth rate. However, the increment in rate is not linear, but tends to saturate at an rf power above 40 W. As a consequence, 40 W is fixed as the working rf power in the following experiments while varying the deposition pressure.

In Fig. 5, the ellipsometric thicknesses of the films grown for 30 min are represented as a function of deposition pressure. Firstly, it can be observed that growth rate shows a maximum at 50 Pa and it decreases significantly for higher pressures. This trend was also found for $HfO_2$ growth in the same HPS system in an $O_2$ atmosphere, with similar deposition conditions [26]. It can be explained by the fact that as pressure rises, the mean free path and the diffusion length of the particles in the plasma become shorter. This prompts a reduction in the flux of sputtered Sc and O atoms that reach the substrate. Znamenskii and Marchenko found a similar dependence for high pressure magnetron sputtering [27] and proposed a diffusion model for the thermalized atoms that is consistent with the results shown here.

Also, Fig. 5 shows that for pressures below 50 Pa, the dielectric film grows at a slower rate. In this case, the explanation can be found with the aid of the plasma optical spectra, which are represented in Fig. 6. In this graph, the presence of neutral and singly-ionized Ar (Ar I and Ar II) in the plasma is observed, regardless of the pressure [28, 29]. However, Sc I peaks [30] are under the detection limit for 25 Pa (only the doublet at 402.0–402.4 nm might be starting to appear) while they are very clearly found for higher pressures. This could be caused by the lower concentration of ions in the plasma which would make the sputter rate decrease.

The ellipsometric results show that these two effects compensate at around 50 Pa and thus the growth rate is at the maximum at that pressure. In fact, the intensity of the Sc I emission can be qualitatively correlated to the growth rate, since the peaks at 50 Pa are the most intense of all spectra. However, direct correlation is difficult, because the confinement of the plasma changes with pressure, so the amount of photons that reach the aperture of the monochromator also changes. In any case, for all pressures the growth rate is reasonable enough in order to obtain nanometric-thick films in the order of minutes. Even for higher pressures, where the growth rates are slower, this could be a positive aspect thanks to the possibility of precisely controlling the thickness of layers.

FTIR spectra of the samples are represented in Fig. 7(a) in the 1200–400 $cm^{-1}$ region. These spectra do not present the broad peak at 600–400 $cm^{-1}$ that was attributed to the Sc\O bond on the thicker layer, probably because of the noticeably lower thickness of these films. In the lowest wavenumber region, only effects of the substrate correction are appreciable. The most relevant feature is the band at around 1050 $cm^{-1}$ that appears more clearly for deposition pressures below 50 Pa. It corresponds with the transverse optical asymmetric stretching vibration mode of the O–Si–O unit, whose tabulated value is 1065 $cm^{-1}$ [31]. The shift of the maximum towards lower wavenumbers might be related to a stress in the silicon oxide film caused by the presence of sub-oxides near the $Si/SiO_x$ interface [32]. For deposition pressures above 75 Pa, this band is noticeably less intense. In Fig. 7(b) the area of the Si–O band is depicted versus the deposition pressure and rf power. In FTIR measurements, the area of a peak is proportional to the concentration of bonds associated with that band [24, 33], in other words, the area of the band is directly related to the film thickness as long as the film densities are similar. Thus, these graphs suggest a thinner $SiO_x$ interlayer for high pressures and for lower rf powers. The $SiO_x$ thickness reduction is a result of the high pressure deposition, which allows fewer O ion to reach the Si surface.

Lastly, the influence of the FGA process and the structure of the MIS capacitors will be analyzed. Fig. 8 shows the spectra of 14 nm and 43 nm thick $ScO_x$ films deposited at 50 Pa of pressure and 40 W of rf power, before and after the FGA at 450 °C. It can be observed that the thicker $ScO_x$ film presents a thicker interfacial $SiO_x$. This fact indicates that the $SiO_x$ grows during $ScO_x$ deposition, since a longer deposition time produces a thicker interfacial $SiO_x$. However, as it is known from growth dynamics [34], this growth is not linear. FTIR results confirm this fact, since the $SiO_x$ peak area is only ~25% smaller for the thinner film, in which the $ScO_x$ is three times thinner. It is also found that the stretching Si–O band at around 1050 $cm^{-1}$ increases after the FGA. The inset in Fig. 8 depicts the increment of the Si–O band area caused by the annealing. From this it is concluded that the FGA process gives rise to a regrowth in the interfacial $SiO_x$ because of the reaction between the $ScO_x$ and the Si substrate, most likely activated by residual $O_2$ present on the film or the RTA furnace.

A TEM image of an Al gated capacitor after the FGA at 450 °C is shown in Fig. 9, where the different layers of the stack can be observed. The scandium oxide film, grown at 50 Pa and 40 W, seems amorphous. This layer is much thinner than the one in Fig. 1, which presents an amorphous structure close to the interface but then a polycrystalline transition. Also, a 3.3±0.3 nm thick $SiO_x$ interlayer appears between the $ScO_x$ and the Si substrate. The thickness of the interface layer is comparable with previous results obtained with $HfO_x$ deposited in similar conditions [35] which also presented an ~3 nm $SiO_2$ interface. The contrast in the upper part of the $ScO_x$ film suggests that the Al gate might be reacting with the dielectric layer, forming an aluminum scandate ($AlScO_x$). This behavior was previously seen by our group when comparing different metal gate electrodes on $ScO_x$ and $GdO_x$ [36].

The AlScO$_x$ might decrease the effective permittivity of the stack [36]. Both effects could influence the C$_{ox}$, as will be seen in the next section.

*3.2. Electrical characterization*

Capacitance and conductance–voltage (C–V and G–V) characteristics at 100 kHz of the sample deposited at 50 Pa and 40 W for 30 min before and after FGA are represented in Fig. 10. Non zero values of the conductance are associated with small-signal energy loss, whose main source is a result from the change in the occupancy of the interfacial traps. Although series resistance must be taken into account, a larger maximum in the conductance would roughly mean a larger density of defects. Fig. 10 shows that after the FGA, both C$_{ox}$ and the conductance are reduced. The decrease in the conductance can be associated to a decrease in the D$_{it}$, as was discussed above. This is probably caused by the passivation of the dangling bonds in the SiO$_x$ by H atoms during the FGA process. Regarding the decrease in the C$_{ox}$, it is observed that the equivalent oxide thickness (EOT) of the MIS capacitor rises from 8.5 nm to 10.0 nm after the FGA. The 1.5 nm increment of the EOT points to the growth of the interfacial layer that is suggested by the FTIR results (Fig. 8). The presence of AlScO$_x$ (Fig. 9), which has a lower permittivity than Sc$_2$O$_3$, would also influence this value. This reaction could be avoided by selecting a different metal gate electrode [36].

Fig. 11 shows the C–V hysteresis curves at 100 kHz of MIS devices with ScO$_x$ deposited for three representative deposition pressures and the same rf power, 40 W. The curves were measured starting in accumulation. For pressures below 50 Pa, the samples present a significantly high flatband voltage shift, ΔV$_{FB}$, of around 0.5 V. This indicates charge trapping in the gate stack defects as a consequence of the gate voltage applied to the device. Nevertheless, layers deposited at higher pressures have much lower shifts. Fig. 12 represents the ΔV$_{FB}$ versus the deposition pressure and rf power. For high pressures, flatband voltage shifts tend to be negligible. Also, higher rf powers produce higher shifts, indicating a lower density of defects in ScO$_x$ films grown with less agressive plasmas. These results highlight the advantages of sputtering at higher pressures, as opposed to lower pressures.

The density of interfacial defects D$_{it}$ was estimated by the conductance method at a frequency of 100 kHz. Results are presented in Fig. 13. From the graph, it can be seen that, as expected, annealing in forming gas consistently reduces the density of defects by a factor of ~2. Also, it is easily deduced that the density of interfacial defects decreases with the deposition pressure. This indicates that higher pressures produce lesser damages to the Si surface, likely during the first stages of the ScO$_x$ growth. This is again due to the reduced reactivity of the plasma in the high pressure regime. Then, it is found that on HPS-grown ScO$_x$ the density of defects is reasonably low and thus remote phonon scattering should be acceptable for device applications. These D$_{it}$ results follow the same trend as previous results of our group which also dealt with ScO$_x$/Si interface [37, 38]. It also corroborates the hysteresis conclusions discussed in the preceding paragraph.

4. **Summary and conclusions**

Structural and electrical characterization of ScO$_x$ thin films grown by HPS have been explored. For long deposition times (2 h), a thick polycrystalline Sc$_2$O$_3$ layer is deposited, with no preferential direction of growth and which reacts with the Si substrate to form a SiO$_x$ interfacial layer. Insight is provided about the kinetics of this interlayer and its relation with the deposition pressure and rf power. The FGA process passivates the SiO$_x$ interface, decreasing the density of defects, but it also prompts a growth of the SiO$_x$ layer or may produce AlScO$_x$ which can affect the EOT. Different gate electrodes should be tested to avoid the metal scandate formation. Sputtering at high pressures (above 100 Pa) reduces the interfacial SiO$_x$ formation, according to FTIR spectra. The devices fabricated with ScO$_x$ deposited at high pressures present lower flatband voltage shifts in the C−V hysteresis curves and lower density of interfacial defects. This shows the advantage of high pressure conditions in reducing the plasma damage of the substrate. On the other hand, growth rates are slower for higher pressures. This fact is not a real drawback since it permits an accurate control of the thickness of thin layers, and the deposition times are in the order of minutes.

**Acknowledgments**

The authors gratefully acknowledge "CAI de Técnicas Físicas," "CAI de Espectroscopía y Espectrometría," "CAI de Difracción de Rayos X" and "CAI de Microscopía Electrónica" of the Complutense University of Madrid for the technical support and "Plataforma de Nanotecnología" of the "Parc Científic" of Barcelona for TEM sample preparation. This work was made possible thanks to the FPU grant AP2007-01157, the FPI grant BES-2011-043798 and the research projects TEC2010-18051 (Spanish "Ministerio de Ciencia e Innovación") and GR58/08 (Complutense University of Madrid).

**References**

[1]     H.H. Tippins, J. Phys. Chem. Solids 27 (1966) 1069.

[2]     R.D. Shannon, J. Appl. Phys. 73 (1993) 348.

[3]     J.A. Kitt, K. Opsomer, M. Popovici, N. Menou, B. Kaczer, X.P. Wang, C. Adelmann, M.A. Pawlak, K. Tomida, A. Rothschild, B. Govoreanu, R. Degraeve, M. Schaekers, M. Zahid, A. Delabie, J. Meersschaut, W. Polspoel, S. Clima, G. Pourtois, W. Knaepen, C. Detavernier, V.V. Afanas'ev, T. Blomberg, D. Pierreux, J. Swerts, P. Fischer, J.W. Maes, D. Manger, W. Vandervorst, T. Conard, A. Franquet, P. Favia, H. Bender, B. Brijs, S. Van Elshocht, M. Jurczak, J. Van Houdt, D.J. Wouters, Microelectron. Eng. 86 (2009) 1789.

[4]     S. Mori, Y.Y. Araki, M. Sato, H. Meguro, H. Tsunoda, E. Kamiya, K. Yoshikawa, N. Arai, E. Sakagami, IEEE Trans. Electron Dev. 43 (1996) 47.

[5] H.M. An, Y.J. Seo, H.D. Kim, K.C. Kim, J.G. Kim, W.J. Cho, J.H. Koh, Y.M. Sung, T.G. Kim, Thin Solid Films 517 (2009) 5589.

[6] M. Roeckerath, J.M.J. Lopes, E. DurgunÖzben, C. Sandow, S. Lenk, T. Heeg, J. Schubert, S. Mantl, Appl. Phys. A 94 (2009) 521.

[7] C. Adelmann, S. Van Elshocht, A. Franquet, T. Conard, O. Richard, H. Bender, P. Lehnen, S. De Gendt, Appl. Phys. Lett. 92 (2008) 112902.

[8] P. Sivasubramani, T.H. Lee, M.J. Kim, J. Kim, B.E. Gnade, R.M. Wallace, L.F. Edge, D.G. Schlom, F.A. Stevie, R. Garcia, Z. Zhu, D.P. Griffis, Appl. Phys. Lett. 89 (2006) 242907.

[9] J.L. Hollenbeck, R.C. Buchanan, J. Mater. Res. 5 (1990) 1058.

[10] D. Schäfer, V. Göpner, R. Wolf, B. Steiger, G. Pfeifer, J. Franke, Proc. SPIE Int. Soc. Opt. Eng. 1782 (1992) 494.

[11] I. Ladany, P.J. Zanzucchi, J.T. Andrews, J. Kane, E. DePiano, Appl. Opt. 25 (1986) 472.

[12] F. Rainer, W.H. Lowdermilk, D. Milam, T.T. Hart, T.L. Lichtenstein, C.K. Carniglia, Appl. Opt. 21 (1982) 3685.

[13] A.C. Jones, T.J. Leedham, H.O. Davies, K.A. Fleeting, P. O'Brien, M.J. Crosbie, P.J. Wright, D.J. Williams, P.A. Lane, Polyhedron 19 (2000) 351.

[14] P.d. Rouffignac, A.P. Yousef, K.H. Kim, R.G. Gordon, Electrochem. Solid-State Lett. 9 (2006) F45.

[15] M. Putkonen, M. Nieminen, J. Niinistö, L. Niinistö, Chem. Mater. 13 (2001) 4701.

[16] M. Toledano-Luque, E. San Andrés, J. Olea, Á. del Prado, I. Mártil, W. Bohne, J. Röhich, E.J. Strub, Mater. Sci. Semicond. Process. 9 (2006) 1020.

[17] E. San Andrés, M. Toledano-Luque, Á. del Prado, M.A. Navacerrada, I. Mártil, G. González-Díaz, W. Bohne, J. Röhrich, E. Strub, J. Vac. Sci. Technol. A 23 (2005) 1523.

[18] S. Dueñas, H. Castán, H. García, E. San Andrés, M. Toledano-Luque, I. Mártil, G. González-Díaz, K. Kukli, T. Uustare, J. Aarik, Semicond. Sci. Technol. 20 (2005) 1044.

[19] J. Mayer, L.A. Giannuzzi, T. Kamino, J. Michael, MRS Bull. 32 (2007) 400.

[20] E.H. Nicollian, J.R. Brews, MOS (Metal Oxide Semiconductor): Physics and Technology, Wiley, New York, 1982.

[21] S.J. Schneider, J.L. Waring, J. Res. Natl. Bur. Stand. A 67A (1963) 19.

[22] M. Toledano-Luque, F.L. Martínez, E. San Andrés, Á. del Prado, I. Mártil, G. GonzálezDíaz, W. Bohne, J. Röhrich, E. Strub, Vacuum 82 (2008) 1391.

[23] M.M. Frank, S. Sayan, S. Dörmann, T.J. Emge, L.S. Wielunski, E. Garfunkel, Y.J. Chabal, Mater. Sci. Eng., B 109 (2004) 6.

[24] A. Hardy, C. Adelmann, S. Van Elshocht, H. Van den Rul, M.K. Van Bael, S. De Gendt, M. D'Olieslaeger, M. Heyns, J.A. Kittl, J. Mullens, Appl. Surf. Sci. 255 (2009) 7812. [25] N.T. McDevitt, W.L. Baun, Spectrochim. Acta 20 (1964) 799.

[26] M. Toledano-Luque, E. San Andrés, Á. del Prado, I. Mártil, M.L. Lucía, G. GonzálezDíaz, F.L. Martínez, W. Bohne, J. Röhrich, E. Strub, J. Appl. Phys. 102 (2007) 044106.

[27] A.G. Znamenskii, V.A. Marchenko, Tech. Phys. 43 (1998) 766.

[28] B. Wende, Z. Phys. 213 (1968) 341.

[29] W.R. Bennett Jr., P.J. Kindlmann, G.N. Mercer, Chem. Lasers: Appl. Opt. Suppl. 2 (1965) 34.

[30] G.A. Martin, J.R. Fuhr, W.L. Wiese, J. Phys. Chem. Ref. Data 17 (Suppl. 3) (1988).

[31] J.F. Fitch, S.S. Kim, G. Lucovsky, J. Vac. Sci. Technol. A 8 (1990) 1871.

[32] D.V. Tsu, J. Vac. Sci. Technol., B 18 (2000) 1976.

[33] S. Miyazaki, H. Nishimura, M. Fukuda, L. Ley, J. Ristein, Appl. Phys. Surf. 113–114 (1997) 585.

[34] S. Wolf, R.N. Tauber, Silicon Processing for the VLSI Era: Vol. 1 — Process Technology, Lattice Press, Sunset Beach, California, 1987.

[35] M. Toledano-Luque, E. San Andrés, Á. del Prado, I. Mártil, M.L. Lucía, G. GonzálezDíaz, F.L. Martínez, W. Bohne, J. Rödrich, E. Strub, J. Appl. Phys. 102 (2007) 044106.

[36] M.A. Pampillón, P.C. Feijoo, E. San Andrés, M. Toledano-Luque, Á. del Prado, A.J. Blázquez, M.L. Lucía, Microelectron. Eng. 88 (2011) 1357.

[37] P.C. Feijoo, Á. del Prado, M. Toledano-Luque, E. San Andrés, M.L. Lucía, J. Appl. Phys. 107 (2010) 084505.

[38] P.C. Feijoo, M. Toledano-Luque, Á. del Prado, E. San Andrés, M.L. Lucía, J.L.G. Fierro, in: Spanish Conference on Electron Devices, 1, 2011, http://dx.doi.org/10.1109/SCED.2011.5744185.

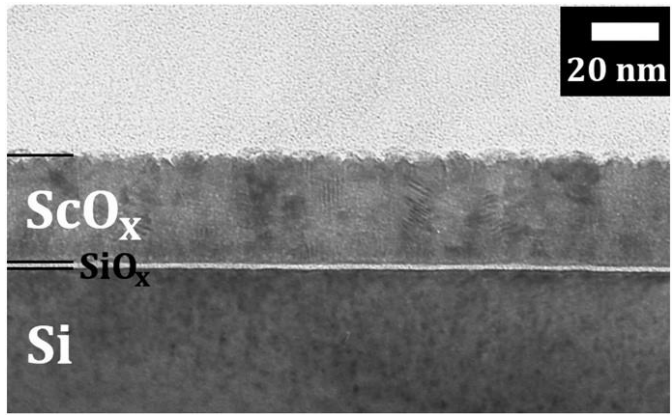

Fig. 1. TEM image of an ~34 nm thick ScO$_x$ film deposited for 2 h on Si at 50 Pa of pure Ar and 40 W of rf power. A 1.7±0.3 nm interfacial layer grows between the dielectric and the substrate.

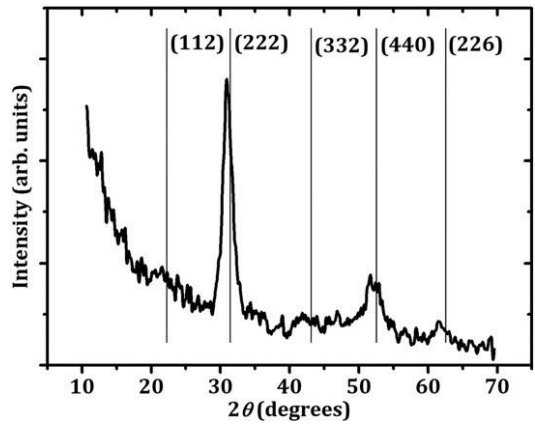

Fig. 2. GIXRD pattern of the thick ScO$_x$ film deposited on Si at 50 Pa, 40 W for 2 h. The most intense peaks of the Sc$_2$O$_3$ cubic bixbyite phase are marked.

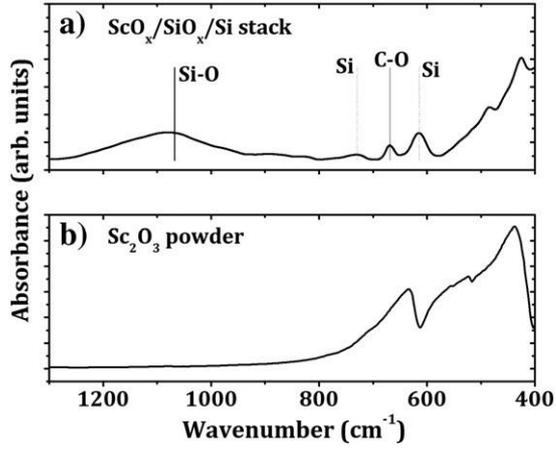

Fig. 3. (a) FTIR spectrum of the sample with a deposition of ScO$_x$ at 50 Pa and 40 W for 2 h. It has been substrate corrected. The increase of absorbance in the 500–400 cm$^{-1}$ region is associated with the Sc\O bonds. (b) FTIR spectrum of Sc$_2$O$_3$ powder.

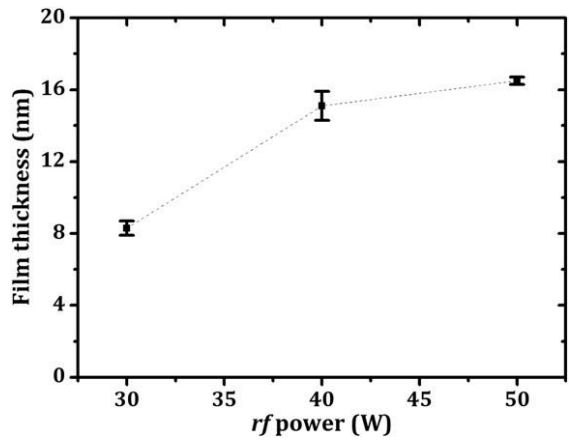

Fig. 4. Film thicknesses of the ScO$_x$ films deposited in pure Ar atmosphere at 50 Pa for 30 min, measured by ellipsometry, as a function of the applied rf power. The rf power was measured with a precision of 1 W.

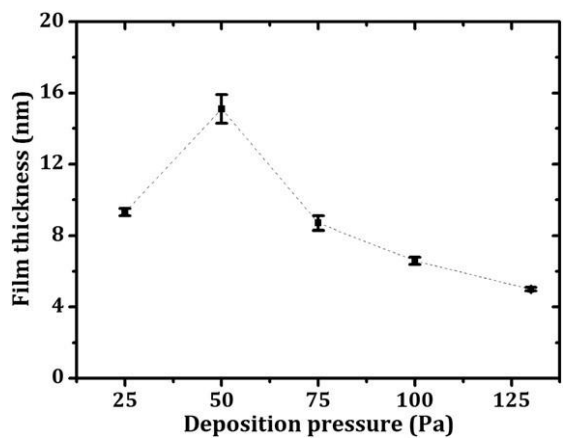

Fig. 5. Film thicknesses of the ScO$_x$ layers for deposition pressures ranging from 25 to 130 Pa of pure Ar, at 40 W of rf power, during 30 min. Measurements were taken by ellipsometry. The deposition pressure was measured with a precision of 1 Pa.

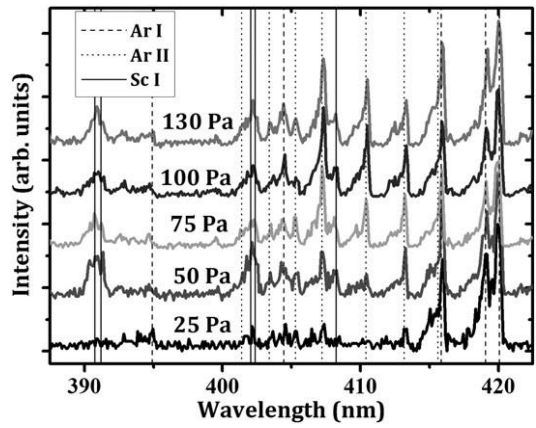

Fig. 6. Optical spectra of $Sc_2O_3$ sputtered at different pressures at 40 W of rf power. The most relevant peaks of Sc I, Ar I and Ar II are indicated by vertical lines. At 25 Pa, the peaks of Sc are under the detection limit.

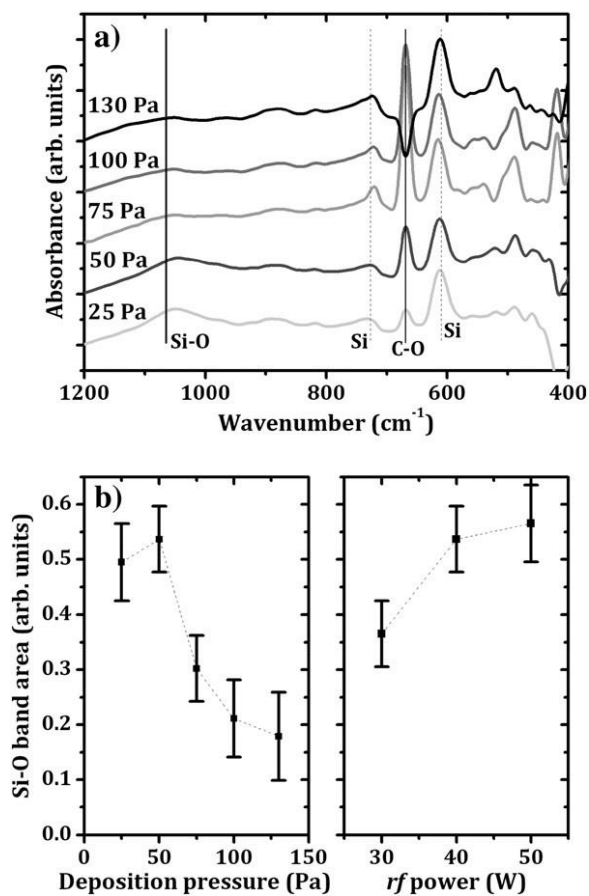

Fig. 7. (a) FTIR spectra of ScO$_x$ layers deposited at pressures ranging from 25 to 130 Pa of pure Ar, at 40 W of rf power, during 30 min. The indicated absorption band (1065 cm$^{-1}$) corresponds to the SiO$_2$ bond stretching. (b) Si–O band area as a function of the deposition pressure at 40 W of rf power (left) and of the deposition rf power at 50 Pa of pressure (right). The rf power was measured with a precision of 1 W and deposition pressure, with 1 Pa. The error bars on the band area were calculated according to the precision of the FTIR spectra.

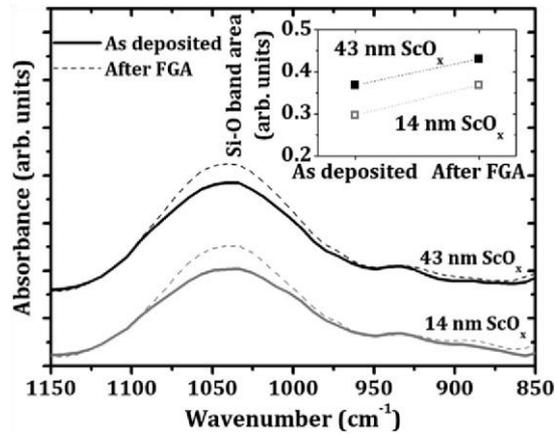

Fig. 8. FTIR spectra of 14 and 43 nm thick ScO$_x$ layers grown at 50 Pa and 40 W before (solid lines) and after the FGA process (dashed lines). In the inset, the area of the Si–O stretching vibration peak (around 1050 cm$^{-1}$) is represented before and after the annealing.

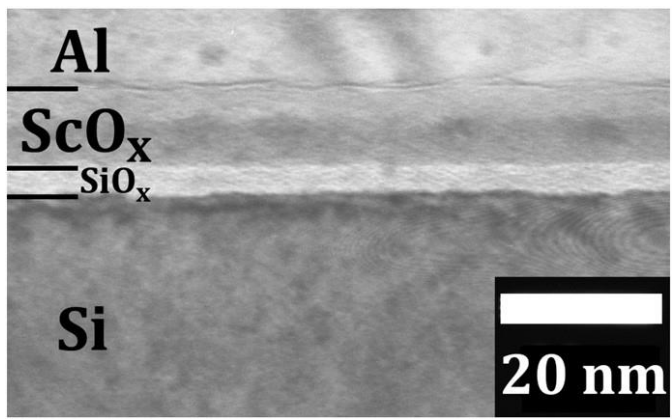

Fig. 9. TEM image of an Al/ScO$_x$/SiO$_x$/Si stack. ScO$_x$ film was deposited at 50 Pa and 40 W. After metallization, the sample went through a FGA process at 450 °C for 20 min.

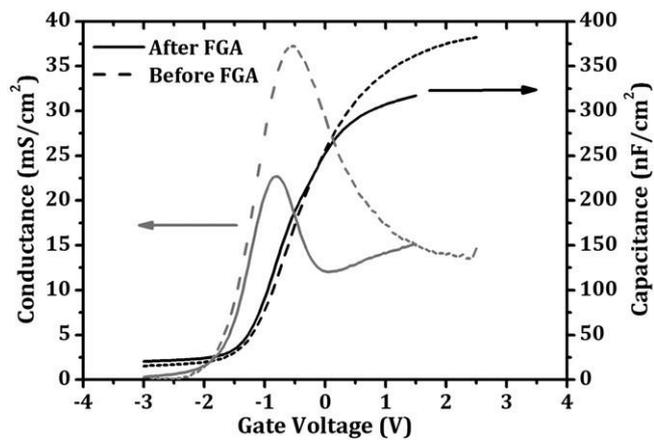

Fig. 10. C–V and G–V curves of the sample deposited at 50 Pa and 40 W for 30 min, before (solid lines) and after the FGA process (dashed lines).

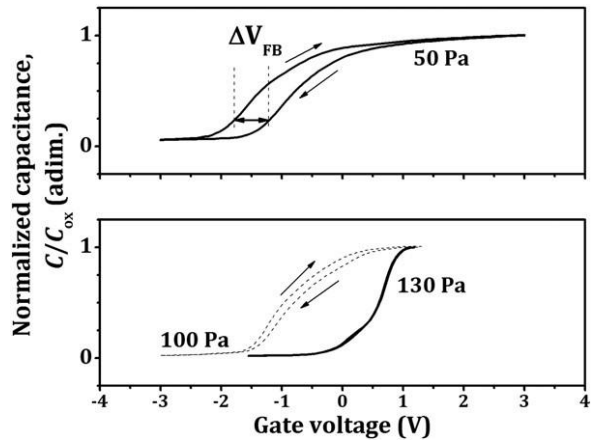

Fig. 11. Representative C−V hysteresis curves of samples deposited at several pressures. For higher pressures, flatband voltage shift is lower.

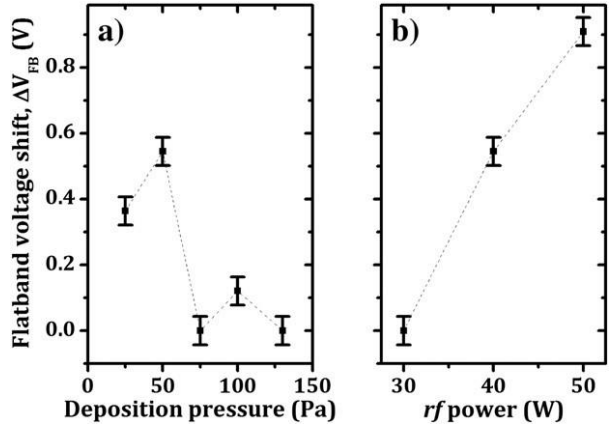

Fig. 12. Flatband voltage shift for Al/ScO$_x$/SiO$_x$/Si MIS devices deposited at (a) different pressures at 40 W of rf power and (b) different rf powers at 50 Pa of pressure. The rf power was measured with a precision of 1 W and deposition pressure, with 1 Pa.

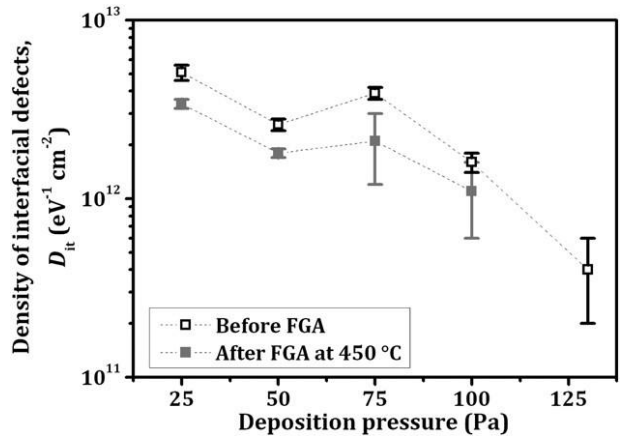

Fig. 13. Density of interfacial defects $D_{it}$ as a function of the deposition pressure, measured by the conductance method. Open symbols represent measurement before FGA and closed symbols, after the FGA at 450 °C. The rf power was measured with a precision of 1 W and deposition pressure, with 1 Pa.